\documentstyle[12pt]{article}

\let\ni=\noindent

\begin{document}

\baselineskip 0.75cm
 
\pagestyle {plain}

\setcounter{page}{1}

\renewcommand{\thefootnote}{\fnsymbol{footnote}}

\newcommand{\CKM}{Cabibbo---Kobayashi---Maskawa }

\pagestyle {plain}

\setcounter{page}{1}

\pagestyle{empty}

~~~

\vspace{0.3cm}

\renewcommand{\thefootnote}{\fnsymbol{footnote}}

{\large\centerline{\bf A four--neutrino texture implying bimaximal flavor 
mixing}}

{\large\centerline{\bf and reduced LSND effect{\footnote{Work supported in part
by the Polish KBN--Grant 2 P03B 052 16 (1999--2000).}}}}

\vspace{0.8cm}

{\centerline {\sc Wojciech Kr\'{o}likowski}}

\vspace{0.8cm}

{\centerline {\it Institute of Theoretical Physics, Warsaw University}}

{\centerline {\it Ho\.{z}a 69,~~PL--00--681 Warszawa, ~Poland}}

\vspace{2.0cm}

{\centerline{\bf Abstract}}

 A four--neutrino effective texture is described, where a sterile neutrino 
mixes nearly maximally with the electron neutrino and so, is responsible for 
the deficit of solar $\nu_e $'s (according to the large--angle MSW solution or 
vacuum solution, of which the former is selected {\it a posteriori}). But, 
while maximal mixing of muon neutrino with tauon neutrino causes the deficit 
of atmospheric $\nu_\mu $'s, the original magnitude of LSND effect is reduced 
by as much as four orders, becoming unobservable.

\vspace{0.2cm}

\ni PACS numbers: 12.15.Ff , 14.60.Pq , 12.15.Hh .

\vspace{3.5cm}

\ni January 2000

\vfill\eject

~~~~
\pagestyle {plain}

\setcounter{page}{1}

\vspace{0.2cm}

 As is well known, in addition to three active neutrinos $ \nu_e,\,\nu_\mu,\,
\nu_\tau $, one sterile neutrino $\nu_s $, at least, is needed to explain in 
terms of neutrino oscillations three neutrino effects: the deficits of solar $
\nu_e $'s and atmospheric $\nu_\mu $'s as well as the possible LSND excess of $
\nu_e $'s in accelerator beam of $\nu_\mu $'s [1]. This is a phenomenological 
reason for introducing sterile neutrinos. From the theoretical viewpoint, 
however, sterile neutrinos may exist in Nature, whether the LSND effect is real
or not.

 In this paper, we describe a four--neutrino effective texture implying 
bimaximal mixing of $\nu_e $ with $\nu_s $ and $\nu_\mu $ with $\nu_\tau $, 
but, at the same time, only a tiny LSND effect, reduced by as much as four 
orders of magnitude in comparison with its original estimation.

 In our texture, the mass matrix for active neutrinos $ \nu_e,\,\nu_\mu,\,
\nu_\tau $ gets the same form $ M = \left( M_{\alpha \beta} \right)\;\,(\alpha,
\,\beta = e,\,\mu,\,\tau)$ as the mass matrix for charged leptons $ e^-,\,\mu^-
,\,\tau^- $ (only the values of parameters are expected to be different). In 
order to operate with an explicit model, we accept in both cases the ansatz [2]

\begin{equation}
\left({M}_{\alpha \beta}\right) = \frac{1}{29} \left(\begin{array}{ccc}
\mu\varepsilon & 2\alpha  & 0 \\ 2\alpha & 4 \mu (80 + \varepsilon)/9 & 
8 \sqrt{3}\,\alpha \\ 0 & 8 \sqrt{3}\,\alpha & 24 \mu (624 + \varepsilon)/25 
\end{array}\right) \;,
\end{equation}

\ni where $ \varepsilon > 0 $, $ \mu > 0 $ and $ \alpha > 0 $ are three 
parameters, taking different values for neutrinos and charged leptons.

 In the case of charged leptons, the ansatz (1) leads for $\alpha \rightarrow 
0 $ to the prediction

\begin{equation}
m_\tau \rightarrow 1776.80\; {\rm MeV}\,,\;\varepsilon \rightarrow 0.172329
\,,\;\mu \rightarrow 85.9924\; {\rm MeV} \,,
\end{equation}

\ni if experimental values of $ m_e $ and $ m_\mu $ are used as an input. In 
fact, the lowest perturbative calculation with respect to $\alpha/\mu $, when 
applied to the eigenvalue equation for the matrix (1), gives in particular [2]

\begin{eqnarray}
m_\tau & = & \frac{6}{125}\left( 351 m_\mu - 136 m_e \right) + 10.2112 
\left(\frac{\alpha}{\mu} \right)^2 \;{\rm MeV} \nonumber \\
& = & \left[ 1776.80 + 10.2112 \left(\frac{\alpha}{\mu} \right)^2\,\right]\;
{\rm MeV} \;.
\end{eqnarray}

\ni When the experimental value $ m_\tau^{\rm exp} = 1777.05^{+0.29}_{-0.26}$
[3] is used, Eq. (3) implies

\begin{equation}
\left(\frac{\alpha}{\mu}\right)^2 = 0.024^{+0.028}_{-0.025} \;,
\end{equation}

\ni what is not inconsistent with $\alpha = 0 $ (then $ M $ becomes diagonal).
Impressive agreement of the prediction for $ m_\tau $ with the experimental
$ m_\tau^{\rm exp} $ is our phenomenological motivation for the use of form (1)
as the lepton mass matrix $ M $. Methodologically, we consider here our form 
(1) of $ M $ as a detailed ansatz, though it can be somehow theoretically 
supported (the interested reader may find some arguments in Appendix to Ref. 
[4]).

 In contrast to the charged--lepton case, where $ \alpha/\mu \ll 1 $ (and so, 
$ M $ is nearly diagonal), we conjecture in the neutrino case that $ \mu/\alpha
\ll 1 $ (and it is small enough to get $ M $ nearly off--diagonal). The reason
is that only in such a situation we can expect nearly maximal neutrino mixing, 
namely of $\nu_\mu $ with $\nu_\tau $ as it is preferably suggested by 
Super--Kamiokande experiments on the deficit of atmospheric $\nu_\mu $'s [5]. 
Then, in order to explain potentially also the deficit of solar $\nu_e $'s [6] 
as well as the possible LSND effect for accelerator $\nu_\mu $'s [7], we accept
the popular hypothesis [1] that in Nature there is a sterile neutrino $\nu_s $ 
which may mix with active neutrinos $\nu_e$, $\nu_\mu $, $\nu_\tau$, dominantly
with $\nu_e $.

 To construct an effective model of four--neutrino texture, we assume that the 
mass matrix for neutrinos $\nu_s $, $\nu_e $, $\nu_\mu $, $\nu_\tau $ has the 
$ 4 \times 4 $ form $ M = \left( M_{\alpha \beta} \right)\;\,(\alpha,\,\beta = 
s,\,e,\,\mu,\,\tau) $, where

\begin{equation}
M_{s\,s} = 0\;,\;M_{s\,e} = \lambda M_{e\,\mu} = M_{e\,s}\;,\;M_{s\,\mu} = 0 =
M_{\mu\,s}\;,\;M_{s\,\tau} = 0 = M_{\tau\,s}
\end{equation}

\ni are seven new matrix elements, while the rest of them are old, as given in 
Eq. (1). Here, the ratio $\lambda \equiv M_{s\,e}/M_{e\,\mu} > 0$ is a neutrino
fourth free parameter. The old neutrino free parameter $\varepsilon $ will be 
put zero (as seen from Eq. (2), even for charged leptons $\varepsilon $ is 
small). Then,

\begin{equation}
M_{e\,e} = 0\;,\;M_{\mu\,\mu} = \frac{4}{9}\,80 \frac{\mu}{29}\;,\;
M_{\tau\,\tau} = \frac{24}{25}\,624 \frac{\mu}{29}\;.
\end{equation}

\ni The ratios 

\begin{equation}
\xi \equiv \frac{M_{\tau\,\tau}}{M_{e\,\mu}} = 299.52\frac{\mu}{\alpha}\;\;,
\;\;\chi \equiv \frac{M_{\mu\,\mu}}{M_{e\,\mu}} = \frac{1}{16.848}\xi 
\end{equation}

\ni are small, when $ \mu/\alpha \ll 1 $ is small enough.

 Now, solving the eigenvalue equation for the $4 \times 4 $ matrix $ M $ in the
first perturbative order with respect to $\xi $, we obtain the following 
neutrino masses:

\begin{eqnarray}
m_0 & = & \!\frac{2\alpha}{29}\left\{-\frac{1}{\sqrt2}\left[49\! + \!\lambda^2
- \sqrt{(49 - \lambda^2)^2 + 4\lambda^2} \right]^{1/2}\!\! + \!\frac{1}{2}
\frac{1}{49}\xi \right\} \simeq \frac{2\alpha}{29} \left[ -\sqrt{\frac{48}{49}}
\lambda + \!\frac{1}{2}\frac{1}{49}\xi \right] \,, \nonumber \\ 
m_1 & = & \!\frac{2\alpha}{29}\left\{\;\,\frac{1}{\sqrt2}\left[ 49\! + \!
\lambda^2 - \sqrt{(49 - \lambda^2)^2 + 4\lambda^2} \right]^{1/2}\!\! + \!
\frac{1}{2}\frac{1}{49}\xi \right\} \simeq \frac{2\alpha}{29} \left[ \sqrt{
\frac{48}{49}} \lambda + \frac{1}{2}\frac{1}{49}\xi \right] \,, \nonumber \\ 
m_2 & = & \!\frac{2\alpha}{29}\left\{-\frac{1}{\sqrt2}\left[ 49\! + \!
\lambda^2 + \sqrt{(49 - \lambda^2)^2 + 4\lambda^2} \right]^{1/2} + \frac{1}{2}
\left(\frac{48}{49} \xi + \chi \right) \right\} \nonumber \\
& \simeq & \frac{2\alpha}{29} \left[-7 + \frac{1}{2}\left(\frac{48}{49}\xi + 
\chi \right)\right] \;, \nonumber \\
m_3 & = & \!\frac{2\alpha}{29}\left\{\;\,\frac{1}{\sqrt2}\left[ 49 + \lambda^2 
+ \sqrt{(49 - \lambda^2)^2 + 4\lambda^2} \right]^{1/2} + \frac{1}{2}\left(
\frac{48}{49} \xi + \chi \right) \right\} \nonumber \\ 
& \simeq & \frac{2\alpha}{29} \left[7 + \frac{1}{2}\left(\frac{48}{49}\xi + 
\chi \right)\right] \;.
\end{eqnarray}

\ni Here, the second step is valid in the linear approximation in $\lambda $,
what requires small $\lambda/49 $, while the former perturbative calculation 
with respect to $\xi $ works for small $\xi/7 $. We can conclude from Eqs. (8) 
that $ m_3 \stackrel{>}{\sim} |m_2| \gg m_1 \stackrel{>}{\sim} |m_0|$.

 The neutrino diagonalizing $ 4\times 4 $ matrix $ U = \left( U_{\alpha\,i}
\right)\;\,(\alpha = s,\,e,\,\mu,\,\tau,\;\, i = 0,\,1,\,2,\,3 )$, such that $ 
U^\dagger M U = {\rm diag}\,\left( m_0,\,m_1,\,m_2,\,m_3 \right)$, gets in the
zero order with respect to $\xi $ and in the linear approximation in $\lambda $
the following form:

\begin{equation}
\left({U}_{\alpha\,i}\right) \simeq \left(\begin{array}{cccc}\frac{1}{\sqrt{2}}
& \frac{1}{\sqrt{2}} & \frac{\lambda}{49\sqrt{2}} & \frac{\lambda}{49\sqrt{2}}
\\ -\frac{\sqrt{48}}{7\sqrt{2}} & \frac{\sqrt{48}}{7\sqrt{2}} & -\frac{1}{7
\sqrt{2}} & \frac{1}{7\sqrt{2}} \\ -\frac{\lambda}{49\sqrt{2}} & -\frac{\lambda
}{49\sqrt{2}} & \frac{1}{\sqrt{2}} & \frac{1}{\sqrt{2}} \\ \frac{1}{7\sqrt{2}} 
& -\frac{1}{7\sqrt{2}} & -\frac{\sqrt{48}}{7\sqrt{2}} & \frac{\sqrt{48}}{7
\sqrt{2}} \end{array}\right) + O(\xi/7) \;.
\end{equation}

\ni Evidently, in this case $\xi/7 $ ought to be smaller than $\lambda/49 $. If
the charged--lepton diagonalizing $ 3\times 3 $ matrix is nearly unit due to 
the small value (4) of $\alpha/\mu $, the lepton counterpart $V = \left( V_{i\,
\alpha}\right)$ of the quark \CKM matrix is approximately equal to $ U^\dagger
= \left( {U^\dagger}_{i\,\alpha} \right) = \left( U^*_{\alpha\,i}\right) $.
Thus, in this approximation, the 
fields

\begin{equation}
\nu_i = \sum_\alpha {V_{i\,\alpha} \nu_\alpha} = \sum_\alpha U^*_{\alpha\,i} 
\nu_{\alpha}
\end{equation}

\ni describe four massive neutrinos $\nu_i \;(i = 0,1,2,3)$ in terms of four 
flavor neutrinos $\nu_\alpha \;(\alpha = s,\,e,\,\mu,\,\tau)$. Hence,

\begin{equation}
\nu_{\alpha} = \sum_i {U_{\alpha\,i} \nu_i }\;\;,\;\;|\nu_{\alpha}\rangle = 
\nu^\dagger_{\alpha}|0 \rangle = \sum_i {U^*_{\alpha\,i}| \nu_i \rangle} \;.
\end{equation}

 Then, the neutrino oscillation probabilities on the energy shell $ E $ read

\begin{eqnarray}
P\left(\nu_\alpha \rightarrow \nu_\beta\right) & = & |\langle \nu_\beta |
e^{i P L}|\nu_\alpha \rangle |^2 \nonumber \\
& = & \delta_{\beta\,\alpha} - 4 \sum_{j>i} U^*_{\beta\,j} U_{\alpha\,j} 
U_{\beta\,i} U^*_{\alpha\,i} \sin^2 x_{ji} \;,
\end{eqnarray}

\ni where  $ L $ denotes the experimental baseline and

\begin{equation}
x_{ji} = 1.27\frac{\Delta m_{ji}^2 L}{E}\;\;,\;\;\Delta m_{ji}^2 = m^2_j
- m^2_i
\end{equation}

\ni with $ \Delta m_{ji}$, $ L $ and $ E $ expressed in eV, km and GeV, 
respectively. In Eq. (12) the eigenvalues of momentum operator $ P $ are $ p_i 
= \sqrt{E^2 - m^2_i} \simeq E - m^2_i/2E$. Evidently, because of real $ M_{
\alpha\,\beta}$ and thus real $ U_{\alpha\,i}$, the possible CP violation is 
here neglected.

From Eqs. (12) and (9) we calculate in the zero perturbative order with respect
to $\xi $ and linear approximation in $\lambda $ the following oscillation 
probabilities:

\begin{eqnarray}
P\left(\nu_e \rightarrow \nu_e\right) & \simeq & 1 - \frac{48^2}{49^2} \sin^2 
x_{10} - \frac{4\cdot 48}{49^2}\sin^2 x_{21} - \frac{1}{49^2} \sin^2 x_{32} \;,
\nonumber \\
P\left(\nu_\mu \rightarrow \nu_\mu \right) & \simeq & 1 - \sin^2 x_{32}\;, 
\nonumber \\ P\left(\nu_\mu \rightarrow \nu_e\right) & \simeq & \frac{1}{49} 
\sin^2 x_{32} \;.
\end{eqnarray}

\ni In the first and third formula (14) we put approximately $\Delta m^2_{20} 
\simeq \Delta m^2_{30} \simeq \Delta m^2_{21} \simeq \Delta m^2_{31}$ due to
Eqs. (8) with $\xi \simeq 0 $ (then, a linear term in $\lambda $ appearing in 
the third formula vanishes). Note from Eqs. (8) that

\begin{equation}
\Delta m^2_{10} \simeq \frac{2}{49}\sqrt{\frac{48}{49}}\left(\frac{2\alpha}{
29}\right)^2 \lambda\,\xi\;\;,\;\;\Delta m^2_{32} \simeq 14 \left(\frac{2
\alpha}{49}\right)^2 \left(\frac{48}{49}\xi + \chi \right)
\end{equation}

\ni for small $ \lambda/49 $ and $\xi/7 $. Here, $\chi = 5.9354 \times 
10^{-2} \xi $.

 If $1.27\,\Delta m^2_{32}\,L_{\rm atm}/E_{\rm atm} = O(1)$ and $\Delta m^2_{
32}\leftrightarrow \Delta m^2_{\rm atm} \sim 3 \times 10^{-3}\;{\rm eV}^2$ [5],
the second formula (14) is able to describe oscillations of atmospheric $
\nu_\mu $'s (dominantly into $\nu_\tau $'s) with maximal amplitude. Then, the 
second Eq. (15) gives the estimate

\begin{equation}
\alpha^2 \xi \sim 4.3 \times 10^{-2}\;{\rm eV}^2 \;.
\end{equation}

\ni Hence, if one assumes reasonably that $\alpha \leq O({\rm 1\;eV})$, one 
gets $\xi \geq O(10^{-2})$.

 On the other hand, if $1.27\,\Delta m^2_{10}\,L_{\rm sol}/E_{\rm sol} = O(1)$ 
and $\Delta m^2_{10} \leftrightarrow \Delta m^2_{\rm sol} \sim 10^{-5}\;{\rm 
eV}^2 $ or $ 10^{-10}\;{\rm eV}^2 $ [6], the first formula (14) can describe 
respectively large--angle MSW oscillations or vacuum oscillations of solar $
\nu_e $'s (dominantly into $\nu_s $'s) with nearly maximal amplitude. In fact, 
it implies

\begin{equation}
P\left(\nu_e \rightarrow \nu_e\right) \simeq 1 - \frac{48^2}{49^2}\sin^2 x_{10}
- \frac{4\cdot 48 + 1}{2\cdot 49^2} \simeq 1 - \frac{48^2}{49^2} \sin^2 x_{10}
\end{equation}

\ni due to $ x_{10} \ll x_{32} \ll x_{21}$. Then, from the first Eq. (15) we 
get the estimate

\begin{equation}
\alpha^2 \lambda\,\xi \sim 5.2 \times 10^{-2}\;{\rm eV}^2\;\;{\rm or}\;\;
5.2 \times 10^{-7}\;{\rm eV}^2 \;,
\end{equation}

\ni respectively.

 Thus, we find from Eqs. (16) and (18) that
 
\begin{equation}
\lambda \sim 1.2 \;\;{\rm or}\;\;1.2 \times 10^{-5}\;,
\end{equation}

\ni respectively. This shows that the matrix element $ M_{s\,e}$ is comparable 
or small {\it versus} $ M_{e\,\mu}$. Evidently, only the first option (related 
to large--angle MSW oscillations of solar $\nu_e $'s) can be compatible with 
the mixing matrix (9) and so, with the oscillation formulae (14) leading to the
nearly maximal mixing of $\nu_s $ with $\nu_e $. In fact, only in this option,
$\xi $ may be smaller than $\lambda $, as required by the form (9) of neutrino
mixing matrix.

 In the case of Chooz experiment searching for oscillations of reactor $\bar{
\nu}_e $'s [8], where it happens that $ 1.27\,\Delta m_{32}^2\,L_{\rm Chooz}/
E_{\rm Chooz} = O(1)$, the first formula (14) leads to

\begin{equation}
P\left(\bar{\nu}_e \rightarrow \bar{\nu}_e\right) \simeq 1 - \frac{1}{49^2} 
\sin^2 x_{32} - \frac{2\cdot 48}{49^2} \simeq 1 
\end{equation}

\ni since $ x_{10} \ll x_{32} \ll x_{21}$. This is consistent with the negative
result of Chooz experiment.

 The third formula (14) implies te existence of $\nu_\mu \rightarrow \nu_e $
neutrino oscillations with the amplitude equal to $ 1/49 \simeq 0.02 $ and the 
mass--squared scale given by $\Delta m_{32}^2$. Such an amplitude is compatible
with the LSND estimation, say, $\sin^2 2 \theta_{\rm LSND} \sim 0.02 $, but the
mass--squared scale $\Delta m_{32}^2 $ --- being equal to the atmospheric $
\Delta m_{\rm atm}^2 \sim 3\times 10^{-3}\;{\rm eV}^2 $ --- is smaller than
the LSND estimation, say, $\Delta m_{\rm LSND}^2 \sim 0.5\;{\rm eV}^2 $ [7] 
roughly by two orders of magnitude.

 In conclusion, our four--neutrino effective texture may describe correctly 
both deficits of solar $\nu_e $'s and atmospheric $\nu_\mu $'s. Then, it 
predicts the existence of a tiny LSND effect of the magnitude reduced by four 
orders in comparison with the original LSND estimation. It is so, because

\begin{equation}
\sin^2 \left( 1.27 \frac{\Delta m_{32}^2\,L_{\rm LSND}}{E_{\rm LSND}} \right)
\sim \sin^2 \left( 1.27 \frac{ 10^{-2}\Delta m_{\rm LSND}^2\,L_{\rm LSND}}{
E_{\rm LSND}} \right) \sim 10^{-4}
\end{equation}

\ni for $ 1.27\,\Delta m_{32}^2\,L_{\rm LSND}/E_{\rm LSND} \sim 1 $. This 
reduced LSND effect would be, therefore, practically unobservable (for original
$ L = L_{\rm LSND}$ and $ E =E_{\rm LSND}$).

 Obviously, the experimental problem of existence of the LSND effect, or of 
another realization of $\nu_\mu \rightarrow \nu_e $ neutrino oscillations, is 
crucial for all discussions about neutrino texture. In particular, a clear 
confirmation of the original LSND effect would exclude our four--neutrino 
effective texture.

 In such a case, the option of three pseudo--Dirac neutrinos might be invoked 
to explain all three neutrino--oscillation effects: the deficits of solar $
\nu_e $'s and atmospheric $\nu_\mu $'s as well as the LSND effect ({\it cf. 
e.g.}, Refs. [4] and [9]). This option involves three natural Majorana sterile
neutrinos mixing nearly maximally with three Majorana active neutrinos, and 
produces three pairs of light mass--neutrino states. It is in contrast to the 
popular see--saw option, where the natural Majorana sterile neutrinos and 
Majorana active neutrinos practically do not mix, and where they produce heavy 
and light mass--neutrino states, respectively. In the see--saw option, small 
masses of the latter states are conditioned by large masses of the former.

\vfill\eject

~~~~
\vspace{0.6cm}

\centerline{\bf References}

\vspace{1.0cm}

{\everypar={\hangindent=0.5truecm}
\parindent=0pt\frenchspacing

{\everypar={\hangindent=0.5truecm}
\parindent=0pt\frenchspacing

~1.~{\it Cf.~~e.g.},~~C.~Giunti,~~Talk at the ICFA/ECFA Workshop,~~Lyon, July 
1999, hep--ph/9910336; and references therein. 

\vspace{0.15cm}

~2.~W. Kr\'{o}likowski, {\it Acta Phys. Pol.} {\bf B 30}, 2631 (1999); and 
references therein.

\vspace{0.15cm}

~3.~Review of Particle Physics, {\it Eur. Phys. J.} {\bf C 3}, 1 (1998). 

\vspace{0.15cm}

~4.~W. Kr\'{o}likowski, "Oscillations of the mixed pseudo--Dirac neutrinos",
{\it Nuovo Cim.} {\bf A} (to appear); and references therein.

\vspace{0.15cm}

~5.~Y. Fukuda {\it et al.} (Super--Kamiokande Collaboration), {\it Phys. Rev. 
Lett.} {\bf 81}, 1562 (1998). 

\vspace{0.15cm}

~6.~{\it Cf. e.g.}, J.N. Bahcall, P.I. Krastev and A.Y. Smirnov, {\it Phys. 
Rev.} {\bf D 58}, 096016 (1998); hep--ph/9905220v2. 

\vspace{0.15cm}

~7.~C.~Athanassopoulos {\it et al.} (LSND Collaboration), {\it Phys. Rev. Lett.
} {\bf 75}, 2650 (1995); {\it Phys. Rev.} {\bf C 54}, 2685 (1996); {\it Phys. 
Rev. Lett.} {\bf 77}, 3082 (1996); {\bf 81}, 1774 (1998).

\vspace{0.15cm}

~8.~M. Appolonio {\it et al.} (Chooz Collaboration), {\it Phys. Lett.} {\bf B 
420}, 397 (1998).

\vspace{0.15cm}

~9.~W. Kr\'{o}likowski, hep--ph/9910308 (to appear in {\it Acta Phys. Pol.} 
{\bf B}); and references therein.

\vfill\eject

\end{document}